\documentclass[aps,showpacs,floatfix,twocolumn,byrevtex,superscriptaddress]{revtex4-1}

\usepackage{amsmath}
\usepackage{amssymb}
\usepackage{amstext}
\usepackage{amsopn}
\usepackage{amsfonts}
\usepackage{amsxtra}
\usepackage[multiple]{footmisc}
\usepackage{bookmark}
\usepackage[english]{babel}
\usepackage{amsmath}
\usepackage{graphicx}
\usepackage{float}
\usepackage{bm}
\usepackage{multirow}
\usepackage{dcolumn}

\usepackage{hyperref}

\begin{document}


\title{Charge-density wave transition in magnetic topological semimetal EuAl$_4$}

\author{R. Yang}
\thanks{These authors contributed equally to this work.}
\affiliation{Key Laboratory of Quantum Materials and Devices of Ministry of Education, School of Physics, Southeast University, Nanjing 211189, China}
\affiliation{1.~Physikalisches Institut Universit$\ddot{a}$t Stuttgart, 70569 Stuttgart, Germany}
\author{C. C. Le}
\thanks{These authors contributed equally to this work.}
\affiliation{RIKEN Interdisciplinary Theoretical and Mathematical Sciences (iTHEMS), Wako, Saitama 351-0198, Japan}
\author{P. Zhu}
\affiliation{ Centre for Quantum Physics, Key Laboratory of Advanced Optoelectronic Quantum Architecture and Measurement (MOE), School of Physics, Beijing Institute of Technology, Beijing 100081, China}
\affiliation{Beijing Key Lab of Nanophotonics and Ultrafine Optoelectronic Systems, Beijing Institute of Technology, Beijing 100081, People's Republic of China}
\author{Z. W. Wang}
\email{zhiweiwang@bit.edu.cn}
\affiliation{ Centre for Quantum Physics, Key Laboratory of Advanced Optoelectronic Quantum Architecture and Measurement (MOE), School of Physics, Beijing Institute of Technology, Beijing 100081, China}
\affiliation{Beijing Key Lab of Nanophotonics and Ultrafine Optoelectronic Systems, Beijing Institute of Technology, Beijing 100081, People's Republic of China}
\affiliation{ Material Science Center, Yangtze Delta Region Academy of Beijing Institute of Technology, Jiaxing 314011, People's Republic of China}
\author{T. Shang}
\affiliation{Key Laboratory of Polar Materials and Devices (MOE), School of Physics and Electronic Science, East China Normal University, Shanghai 200241, China}
\author{Y. M. Dai}
\email{ymdai@nju.edu.cn}
\affiliation{National Laboratory of Solid State Microstructures and Department of Physics, Nanjing University, Nanjing 210093, China}
\affiliation{Collaborative Innovation Center of Advanced Microstructures, Nanjing University, Nanjing 210093, China}
\author{J. P. Hu}
\affiliation{Institute of Physics, Chinese Academy of Sciences, Beijing 100190, China}
\author{M. Dressel}
\email{dressel@pi1.physik.uni-stuttgart.de}
\affiliation{1.~Physikalisches Institut Universit$\ddot{a}$t Stuttgart, 70569 Stuttgart, Germany}
\date{\today}

%

\begin{abstract}
The interplay among topology, charge-density wave (CDW), and magnetism can give rise to a plethora of exotic quantum phenomena. Recently, a group of magnetic topological semimetals with tetragonal lattices and CDW order were found to exhibit anomalous magnetic instability, helical spin ordering, and the presence of skyrmions. However, the underlying mechanism responsible for these observations remains unclear. Here, we conducted a comprehensive investigation into the impact of CDW on the topological and magnetic properties of EuAl$_4$ using optical spectroscopy and the first-principles calculations. Through optical spectroscopy, we observed a partial gap (60~meV) on the Fermi surface and an enhanced mid-infrared absorption around 0.4~eV after the CDW transition. Magneto-optical spectroscopy and the first-principles calculations proved that, by affecting the band structure, the CDW order frustrates the antiferromagnetic interactions but strengthened the ferromagnetic ones, which can destabilize the magnetism. With lower symmetry in the CDW ordered state, carriers from the Weyl bands will mediate the anisotropic magnetic interactions promoting the formation of chiral spin textures. Conversely, without the CDW order, the counterpart EuGa$_4$ shows robust collinear antiferromagnetic order. Our findings uncover the pivotal role played by CDW order in arousing intricate magnetism in topological materials and provide valuable insights into controlling topological and magnetic properties through the manipulation of CDW orders.
\end{abstract}


\maketitle
%
\section{Introduction}
The interplay among topology, many-body effects, and magnetism represent a cutting-edge research frontier in condensed matter physics~\cite{Nagaosa2013, He2022, Bernevig2022}.
The coupling between topology and charge-density wave (CDW) can give rise to an axion insulator state~\cite{Shi2021}, while the entanglement of magnetism and topology leads to intriguing phenomena such as quantum anomalous Hall effects and skyrmions~\cite{Bernevig2022}.
Rare-earth intermetallic compounds, in which the local moments of rare-earth atoms interact with the itinerant carriers from topological bands, providing an ideal platform for studying and manipulating novel topological physics~\cite{Sanchez2020, Ma2020, Hirschberger2016, Manna2018, Cheng2021}.
Recently, the CDW transition was observed in a group of rare-earth magnetic topological semimetals with tetragonal lattices.
The charge modulation affects the band topology as well as the magnetism, leading to unexpected magnetic and topological properties including multi-q magnetic order, helical spin order and even skyrmions~\cite{Takagi2022, Shang2021,Yasui2020, Khanh2020, Moya2022, Lei2019, Lei2021}.
Nevertheless, the underlying mechanism arising from the interplay among topology, magnetism, and CDW in one system remain unclear.

Among the rare-earth intermetallic compounds, the binary compounds EuM$_4$ (M= Al, Ga) family provide an unique arena for investigating the intricate interplay between CDW, topology, and magnetism due to their stoichiometric composition and relatively simple band and lattice structures ($I4/mmm$)~\cite{Takagi2022, Shang2021,Ramakrishnan2022}.
In EuM$_4$, magnetism originates from the local $4f$ electrons of the Eu$^{2+}$ atoms that are sandwiched between M$_4$ layers. The magnetic interactions are mediated either by itinerant carriers in M$_4$ layers through Ruderman-Kittel-Kasuya-Yosida (RKKY) interaction (antiferromagnetic, AFM) or by local excitations (ferromagnetic, FM), giving rise to A-type AFM order at low temperatures ($T$s)~\cite{Nakamura2015}.
For EuAl$_4$, a CDW transition occurs at 145~K and a consecutive of intricate magnetic orders with non-coplanar spin texture forms below 15.6~K~\cite{Stavinoha2018}.
The magnetism in EuAl$_4$ is highly susceptible and can be easily disrupted by a weak magnetic field ($<$2~T), resulting in unexpected helical spin order and skyrmions in such centrosymmetric lattice~\cite{Takagi2022}.
In contrast, the counterpart EuGa$_4$, which lacks CDW, exhibits a highly stable coplanar magnetic order up to 7~T~\cite{Zhang2021}.
Given that EuAl$_4$ and EuGa$_4$ have the same lattice structure and similar band structures~\cite{Nakamura2015}, it is evident that the presence of CDW order plays a decisive role in influencing the magnetism as well as the topology of EuAl$_4$.
Previous studies on both EuAl$_4$~\cite{Kobata2016} and EuGa$_4$~\cite{Lei2023} have observed Dirac nodal lines near the Fermi level contributing to itinerant carriers that mediating the magnetic interactions.
Furthermore, recent experimental~\cite{Shang2021, Ramakrishnan2022} and theoretical~\cite{Wang2023, Kaneko2021} investigations indicated a small nesting vector connecting the Dirac-like bands.
Nevertheless, the nature of the CDW gap as well as how the CDW transition affects the interplay between topology and magnetism still remain enigmatic.

Optical spectroscopy is a powerful technique for investigating charge excitations in solids~\cite{Dressel2002, Corasaniti2019, Huang2013, Hu2008}, particularly under magnetic fields, providing insights into the underlying charge-spin interactions.
In this work, we conducted a comparative investigation between EuAl$_4$ and EuGa$_4$ to elucidate the effects induced by CDW transition by means of optical spectroscopy measurements and first-principles calculations.
In comparison with its isostructural counterpart EuGa$_4$, we firstly identified the CDW gap (60~meV) in EuAl$_4$. The CDW order not only partially eliminates the Fermi surface but also affects the high-energy excitations around 0.4~eV. Further magneto-optical measurements and theoretical calculations prove that these behavior modulate the magnetic interactions, destabilizing the magnetism. In combination with the broken symmetry induced by lattice distortion, enhanced FM responses, and the partial gap on the Fermi surface, the anisotropic magnetic interactions will promote the noncoplanar spin textures.

\section{Results}
%
%
\begin{figure*}[tb]
\centerline{
\includegraphics[width=1.8\columnwidth]{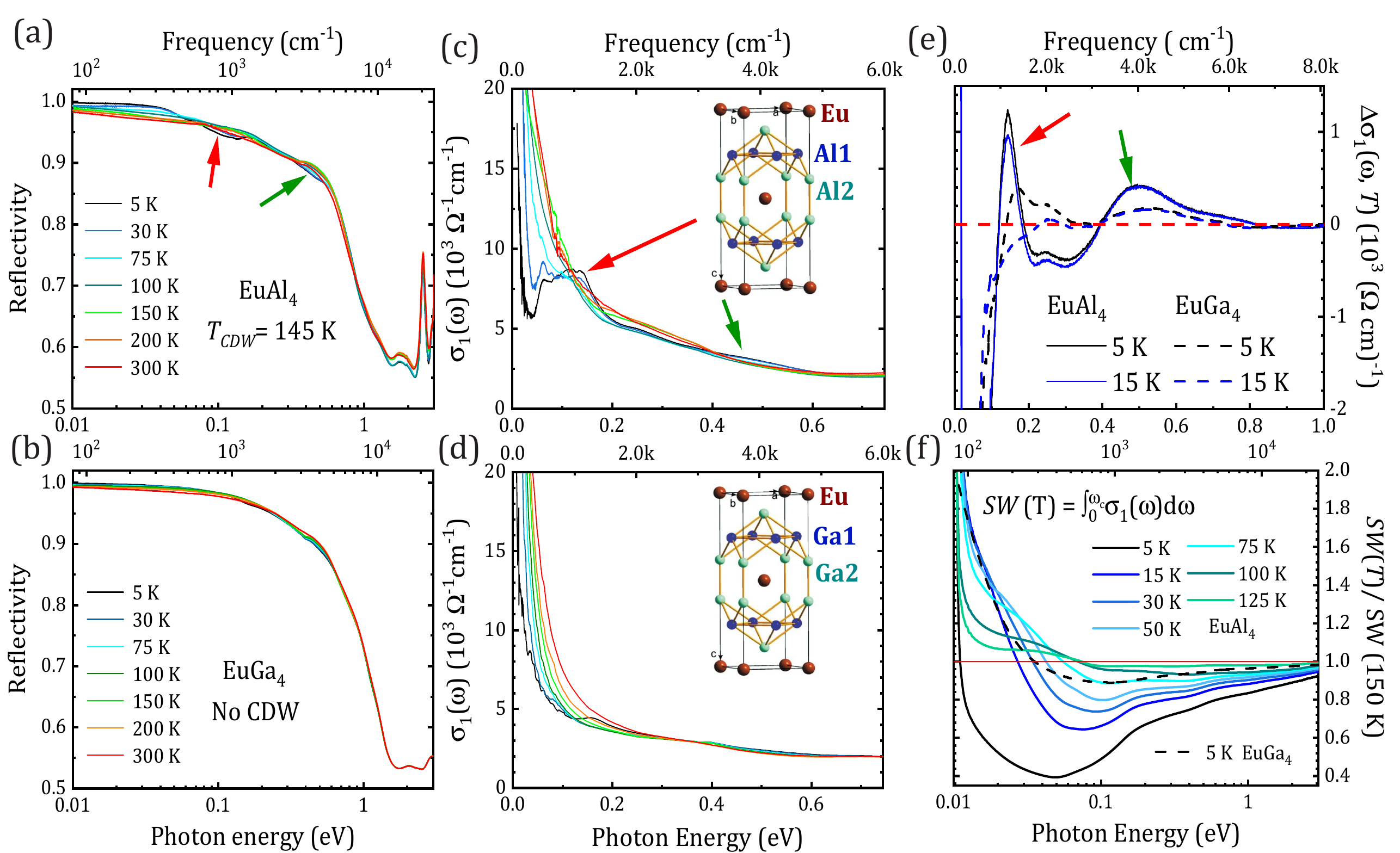}
}
\caption{\textbf{Optical spectroscopy of EuAl$_4$ and EuGa$_4$.}  Temperature-dependent reflectivity of (a) EuAl$_4$ and (b) EuGa$_4$ from 80 to 24\,000~cm$^{-1}$ with the light polarized in the $ab$-plane. In panel (a), the reflectivity changes in EuAl$_4$ below $T_{\rm CDW}=145$~K are highlighted by red and green arrows. (c-d) Temperature evolution of the real part $\sigma_1(\omega;T)$ of the optical conductivity spectra of (c) EuAl$_4$ and (d) EuGa$_4$ below 6\,000~cm$^{-1}$ (0.74~eV); the emergent absorption features after the CDW transition are indicated by red and green arrows in panel (c).
The insets display the lattice structure of EuAl$_4$ and EuGa$_4$, which contain two inequivalent Al/Ga atoms. (e)~Difference spectra of $\sigma_1(\omega)$ below  the CDW transition calculated through $\Delta\sigma_1(\omega, T)= \sigma_1(\omega, T)-\sigma_1(\omega, 150~{\rm K})$. The newly developed absorption is indicated by the colored arrows. The solid and dashed lines represent the EuAl$_4$ and EuGa$_4$ data, respectively. (f) The integrated spectral weight ($SW$) of EuAl$_4$ according to Eq.~(\ref{integral SW}), normalized by the $SW$ at $T=150$~K, is plotted for $T< T_{\rm CDW}$ as a function of the cutoff frequency $\omega_c$; the $SW$ redistribution takes place in a broad energy range.}
\label{fig:optical spectroscopy}
\end{figure*}

\subsection{Optical spectroscopy}
Figures~\ref{fig:optical spectroscopy} (a) and (b) display EuAl$_4$'s and EuGa$_4$'s temperature ($T$)-dependent reflectivity $R(\omega)$ spanning from the far-infrared (FIR) to ultraviolet (UV) range, respectively; details of the measurements are given in the supplementary materials (SM)~\cite{Supplementary}.
The high reflectivities ($>0.9$) that gradually increase with decreasing $T$ and a pronounced plasma edge around 0.9~eV evidence their metallic nature.
When $T < T_{\rm CDW}$, EuAl$_4$'s $R(\omega)$ is suppressed around 0.1 (red arrow), 0.5 (green arrow), and 1.5~eV, indicating emergent absorptions after the CDW transition~\cite{Corasaniti2019, Hu2008, Dressel2002}, while no additional structure develops in EuGa$_4$'s $R(\omega)$ from 300 to 5~K.
Based on the Kramers-Kronig analysis, the optical conductivity was derived from $R(\omega)$ (see detail in the SM~\cite{Supplementary}).
For both compounds, the optical conductivities are plotted in Figs.\ref{fig:optical spectroscopy}(c) and (d), respectively. The real part of the optical conductivity, $\sigma_1 (\omega)$, which reflects the joint density of states~\cite{Dressel2002}, displays intraband zero-energy modes (Drude peak) that roll off with a characteristic width, which represents the scattering of itinerant carriers.
With increasing photon energy, the Drude peaks gradually develops into a series of interband absorptions (Lorentz peaks).
Across the CDW transition, in EuAl$_4$'s $\sigma_1 (\omega)$ (Fig.\ref{fig:optical spectroscopy}(c)), we encounter a great depletion of intraband responses with emergent absorption peaks around 0.1 and 0.4~eV,  signaling the formation of CDW gap on the Fermi surface and a band reconstruction at high-energy range~\cite{Corasaniti2019, Hu2008}.
Since the intraband responses persist to 5~K, the Fermi surface is only partially gapped.
At 5~K, the newly formed absorptions around 0.1~eV give rise to a plateau-like structure, which cannot be ascribed to a single Lorentz peak.
In contrast, without the CDW transition, EuGa$_4$'s $\sigma_1 (\omega)$ is remarkably different: only a continuous narrowing of the Drude peak is observed down to 15~K(Fig.\ref{fig:optical spectroscopy}(c)), reflecting diminishing scattering rate upon cooling.
When entering into AFM state, with the suppressed spin fluctuations, the Drude peak exhibits a remarkable narrowing and a small bump emerges at 0.17~eV.
In Fig.~\ref{fig:optical spectroscopy}e, the difference spectra $\Delta \sigma_1 (\omega)$ respect to 150~K were calculated for both samples.
They further show the affections caused by CDW transition~\cite{Yang2019, Uykur2022}.
For EuAl$_4$, the negative $\Delta \sigma_1 (\omega)$ at low energy range and two remarkable absorption peaks around 0.1 (red arrow) and 0.4~eV (green arrow) deliver the message that besides the partially gap on the Fermi surface, the high-energy excitations are also affected (green arrow).
However, the changes in EuGa$_4$'s $\sigma_1 (\omega)$ are no that dramatic and mainly come from the temperature effects.

To elucidate the $SW$ redistribution caused by the CDW transition in EuAl$_4$, we calculated the integrated $SW$ of the measured $\sigma_1 (\omega)$ up to the cutoff frequency ($\omega_c$), which is given by~\cite{Dressel2002, Corasaniti2020}:
\begin{equation}
    SW (\omega_c;T)=\frac{Z_0}{\pi}\int^{\omega_c}_{0}\sigma_1 (\omega';T)d\omega',
\label{integral SW}
\end{equation}
expressed in units of cm$^{-2}$ ($Z_0=$377~$\Omega$ being the impedance of vacuum). Such model-independent value is related to the carriers (normalized to their effective mass) contributing to the optical excitations up to $\omega_c$ and reflects the evolution of the band structures at various temperatures.
In the limit $\omega\rightarrow \infty$, the $SW$ is expected to converge to a constant value, satisfying the optical $f$-sum rule~\cite{Dressel2002}.
To show the affection from CDW transition, we calculate the ratio $SW (\omega_c;T)/SW (\omega_c;150~{\rm K})$, which underscore the energy range of the $SW$ reshuffling as a function of $T$ with respect to 150~K, which is slightly above $T_{\rm CDW}$
\footnote{If there is a transfer of $SW$ from high to low energies, the $SW$ ratio will exceed 1 at low energies and then smoothly approach 1 upon increasing $\omega$ until the full energy scale of the low-energy resonance is reached. If there is a transfer of SW from low to high energies, the $SW$ ratio will fall below 1 until the total energy scale of $SW$ transfer is reached. The latter case suggests some depletion of density of states (DOS), as it would occur when an electronic bands reconstruction happens.}.
The results presented in Fig.\ref{fig:optical spectroscopy}f display a twofold $SW$ reshuffling to low and high energy ranges for both samples.
In the low-energy range, the narrowing of the Drude peak gives rise to the accumulation of $SW$ in a very small FIR energy range and a ratio above 1.
In EuAl$_4$, the suppressed intraband responses caused by the opening of CDW gap result in a ratio $SW (\omega_c;T)/SW (\omega_c;150~{\rm K})$ far below 1, and its minimum corresponds to the energy scale of the single-particle gap excitation within the electronic structure, based on which the CDW gap is estimated to be 66~meV (5~K)~\cite{Corasaniti2019}.
Even though the $SW$ starts to recover above the gap, it is not fully retrieved until 3~eV, which is the highest energy for our measurements, indicating a very broad energy range for the $SW$ reshuffling.
Such behavior has been widely observed in LiV$_2$O$_4$~\cite{Jonsson2007}, iridates~\cite{Wang2021}, and cuprates~\cite{Basov2005}, reflecting a strong correlation effect, which is further confirmed by estimating the renormalization of electronic kinetic energy (see section III of the $SM$~\cite{Supplementary})~\cite{Xu2020}.
Thus, the $SW$ analysis also confirms the partial gap and the enhanced high-energy excitations after the CDW transition.
Such tendency is additionally enhanced by the correlation effect.
However, without the CDW order, EuGa$_4$'s $SW$ is just marginally suppressed and shifted to very high-energy range, primarily attributed to the correlation effect.

%
\begin{figure}[tb]
\centerline{
\includegraphics[width=1.0\columnwidth]{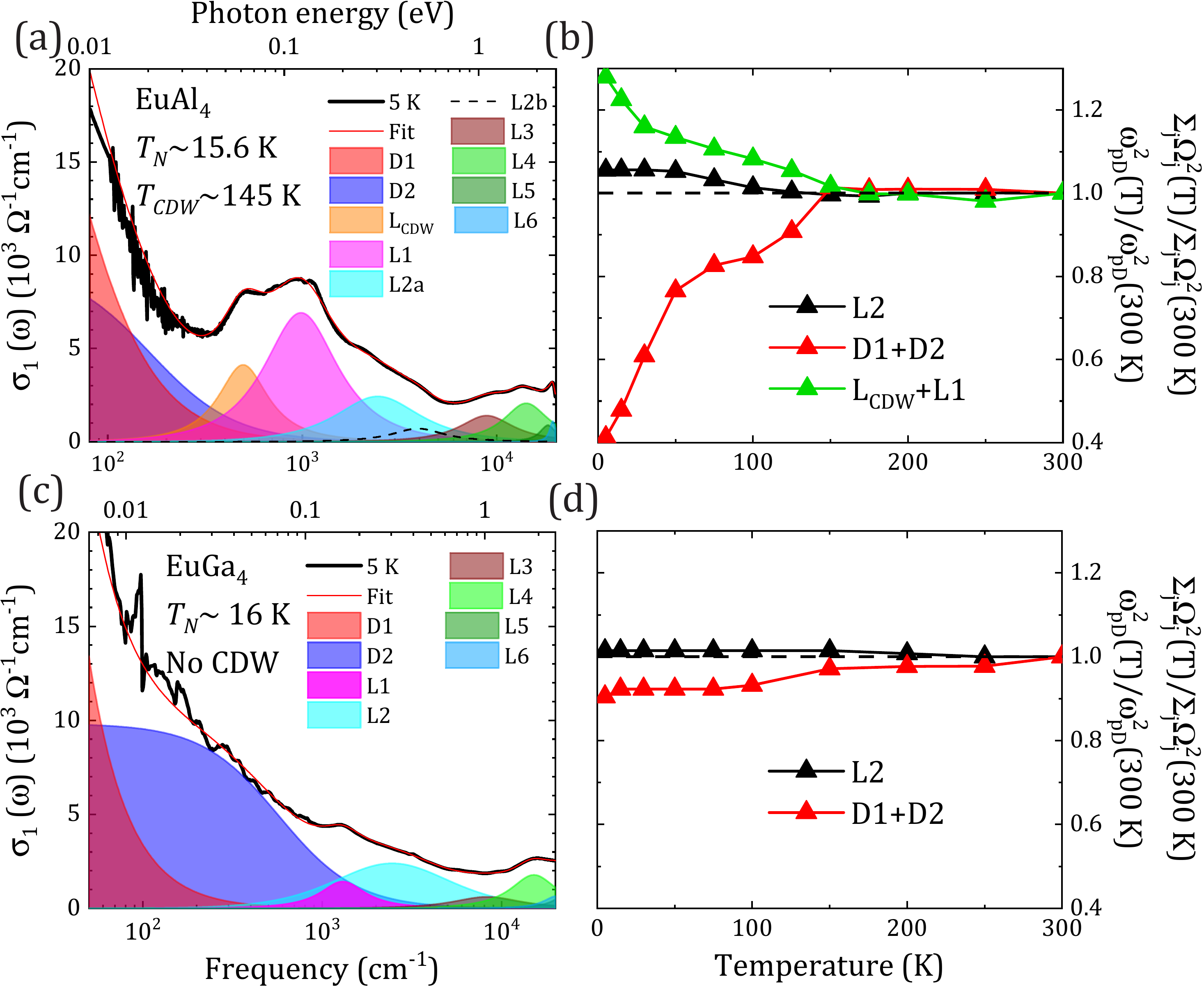}
}
\caption{\textbf{Drude(D)-Lorentz(L) fit to the optical conductivity.} The fits to EuAl$_4$'s (a) and EuGa$_4$'s (c) optical conductivity at 5~K; the optical conductivities can be decomposed into the narrow and broad Drude components, as well as several Lorentz oscillators. (b) and (d) are $T$-dependent spectral weight of the Drude components ($\omega^{2}_{pD}=\omega^{2}_{pD1}+\omega^{2}_{pD2}$) and the Lorentz components ($\sum_{j}\Omega^2$) normalized to the value at 300 K. Refer to section III of the SM~\cite{Supplementary} for more details.}
\label{fig:D-L fit}
\end{figure}

\subsection{Drude(D)-Lorentz(L) fit}
With the goal to quantitatively describe the electrodynamic response across the CDW transition, the $\sigma_1 (\omega)$ of EuAl$_4$ and EuGa$_4$  were fit within the common Drude-Lorentz phenomenological approach (we refer to section III of  Ref.~\cite{Supplementary} for details of the fit procedure).
The resulting fits with their constituent components are displayed in Figs.~\ref{fig:D-L fit} (a) and (c) and Fig. S3 in the SM~\cite{Supplementary}.
At high $T$s, both samples' $\sigma_1 (\omega)$ can be described by two Drude (D) components with different width (scattering rate) and the same number of Lorentzian (L) oscillators , signaling their similar band structures (see Fig. S3 and the discussion in section III of the SM~\cite{Supplementary} for detail).
However, after the CDW transition, EuAl$_4$'s two Drude components are significantly suppressed, giving way to a newly formed Lorentz peak around 60~meV coming from the CDW gap that partially opens on the Fermi surface Figs.~\ref{fig:D-L fit} (a).
This is best illustrated by comparing Fig.~\ref{fig:D-L fit}(a) with (c).
Through an analysis of the SW distribution of each intraband and interband responses (the SW was defined as squared plasma frequency ($\omega_{pD}^{2}$) or oscillator strength ($\Omega_{j}^{2}$) of each fit component), we notice that, in EuAl$_4$, above $T_{CDW}$, the SW of each component does not show discernable change, while, in the CDW ordered state, the SW of intraband responses is significantly suppressed by almost 60\% (5~K) and transferred to either CDW absorptions or mid-infrared (MIR) interband transitions around 0.4~eV (Fig.~\ref{fig:D-L fit} (c)).
In contrast, the SW of the MIR absorption in EuGa$_4$ shows almost no $T$ dependence (Fig.~\ref{fig:D-L fit} (d)), a marginal suppression of the Drude components can be attributed to the correlation effects.

%
\begin{figure}[tb]
\centerline{
\includegraphics[width=1.05\columnwidth]{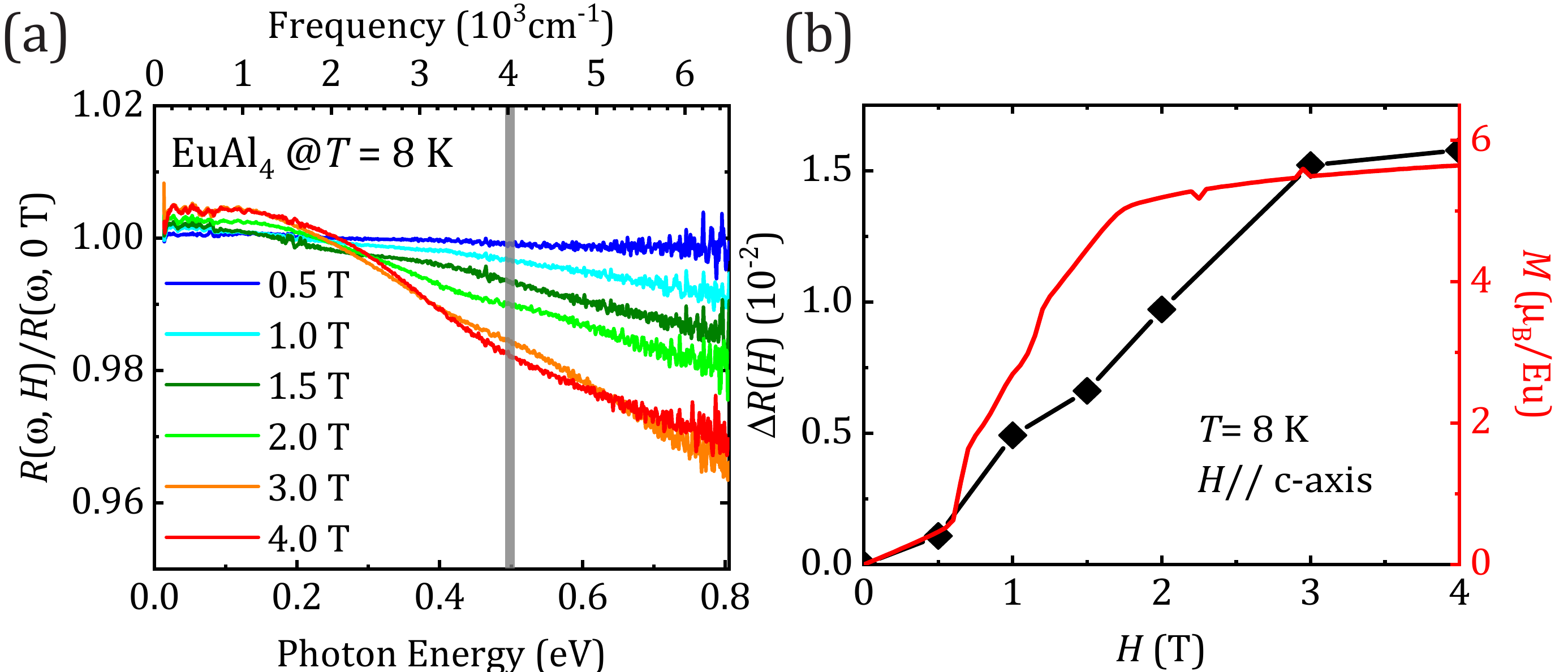}
}
\caption{\textbf{Magneto-optical response of EuAl$_4$.}(a) Magneto-optical reflectivity spectra $R(\omega, H)$ measured under out-of-plane fields $H$ from 0 to 4~T and normalized to the zero-field reflectivity $R(\omega, 0~{\rm T})$. (b)~Comparison between the field-dependent reflectivity change $\Delta R(H) = [1-R(H)/R(0 T)]$ at 0.5~eV (the gray bar in (a)) and the magnetization (red). The overall field variation is rather similar except some deviations most likely due to more complicated mechanisms for spin reorientation.}
\label{fig:MO}
\end{figure}

\subsection{Magneto-optical responses}
Up to now, we have learnt that, besides the CDW gap, the other difference between EuAl$_4$ and EuGa$_4$ is the enhanced MIR absorptions.
To examine the relation between MIR absorptions and magnetism, we further measured the magneto-optical spectra with an \emph{in-situ} magnetic field along the c-axis ($H\parallel c$).
The results shown in Fig.~\ref{fig:MO}a exhibit no discernable change in reflectivity below 0.5~T.
For $H>$0.5~T, however the low-energy reflectance increases slightly and the reflectivity above 0.25~eV is continuously suppressed; this behavior saturates at 3~T.
In light of the $T$-dependent $R(\omega)$ (Fig.~\ref{fig:optical spectroscopy}a and Fig. S2b in the SM~\cite{Supplementary}), the suppression in reflectivity across the CDW transition stems from the absorptions of the CDW gap in the FIR range and enhanced MIR optical responses~\cite{Chen2017}.
Fig.~\ref{fig:MO}b displays the change in reflectivity under magnetic field $\Delta R (H)$ measured at 0.5~eV ~\footnote{In Fig. S2b of SM, the dip of the MIR absorption in the reflectivity ratio $R(\omega, 5~{\rm K})/R(\omega, 150~K)$ resides around 0.5~eV; thus we use the data at this point to trace the evolution of the MIR absorptions under magnetic field.}. 
The similar tendency between the MIR absorptions and the magnetization until when the Eu moments are fully aligned signals an intimate relation between the MIR absorptions and the FM responses in EuAl$_4$.
On the other side, the slightly enhanced $R(\omega)$ below 0.25~eV reflects the degenerated CDW gap and the restoration of metallicity while aligning the local moments, suggesting a coupling between local moments and itinerant carriers.
%
\begin{figure*}[tb]
\centerline{
\includegraphics[width=1.8\columnwidth]{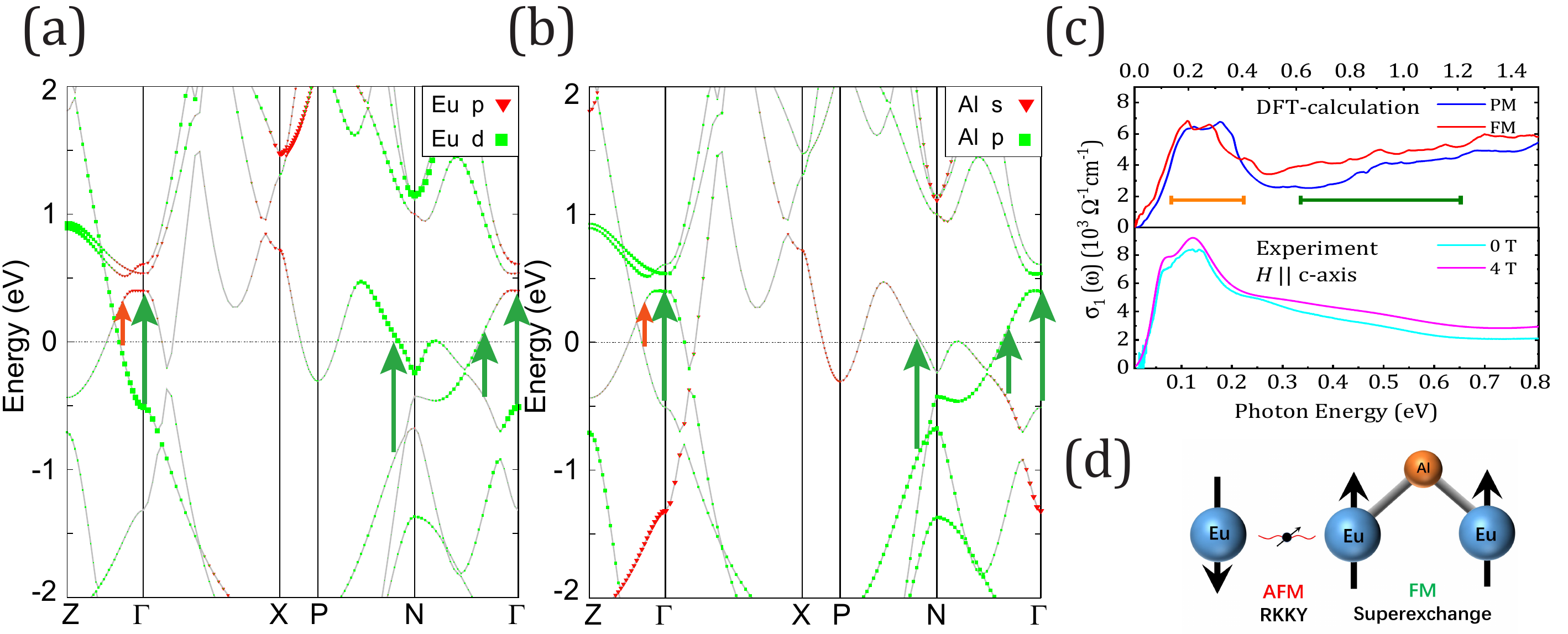}
}
\caption{\textbf{The first-principles calculations.}(a) and (b) display the band structures of EuAl$_4$ in PM state without the CDW order; the contribution from  Eu $5d$ and Al $3p$ orbitals are highlighted in (a) and (b), separately. The upper and lower panels of (c) show the calculated and the measured interband optical conductivities. Blue and red lines in upper panel correspond to the absorptions in PM and FM (Eu's moments along the c axis) states, and turquoise and pink lines are conductivities measured 0 and 4~T (Eu's moments are fully aligned to the c-axis). Orange and green segments represent the interband transitions denoted by the same colored arrows in (a) and (b). (d) is the sketch of magnetic interactions in EuAl$_4$.}
\label{fig:DFT}
\end{figure*}

\subsection{The first-principles calculations}
Next we calculate the band structure and simulate the optical conductivity based on the density functional theory (DFT) to trace the origin of MIR absorptions and their relation to magnetism.
The band structures in the paramagnetic (PM) state (without CDW order) are shown in Figs.~\ref{fig:DFT}a and b.
At low energies, several bands cross the Fermi level giving rise to hole and electron pockets; along the $\Gamma-Z$ direction of the Brillouin zone, two linear bands cross each other generating a Dirac cone, in line with the Dirac semimetal nature of EuAl$_4$.
From the perspective of orbital composition, the bands near the Fermi level are dominated by Eu $5d$ and Al $3p$ orbitals as illustrated by colors in Figs.~\ref{fig:DFT}a and b.
The overlap of these orbitals in several bands indicates considerable hybridizations between them~\footnote{Eu $5d$ orbitals mainly hybridize with Al2 orbital, as Al2 is closer to Eu atoms, see the inset of Fig.~\ref{fig:optical spectroscopy}c and Fig. S4 in SM}.
Such $pd$ hybridization and excitations between Eu $5d$ and Al $3p$ orbitals were designated as the bridge delivering FM exchange interactions~\cite{Nomoto2020, Anderson1950}.
The lattice distortion within the Al$_{4}^{2+}$ layers will inevitably change the distance between Eu and Al atoms~\cite{Ramakrishnan2022}, thereby affecting the FM interactions.

The overall band-structure calculations lay the foundation for obtaining the interband components of the optical conductivity.
Figure~\ref{fig:DFT}c displays the calculated $\sigma_1(\omega)$ of EuAl$_4$ (upper panel) compared with the measured spectrum(lower panel)~\cite{Yang2022}.
To mimic the magnetic field effect, in the upper panel, we calculated the conductivity in both PM and field-forced FM states (the band structure with Eu moments along the $c$-axis can be seen in Fig. S6a of SM~\cite{Supplementary}).
In the lower panel of Fig.~\ref{fig:DFT}c, by fit the measured reflectivity at 4~T , the interband $\sigma_1(\omega)$ at 4~T was reproduced (see SM~\cite{Supplementary} for detail).
Figure~\ref{fig:DFT}c demonstrates that the theoretical results can well reproduce the observation in either line shape and magnetic dependency.
The difference in energy is due to the band renormalization caused by the correlation effects which were not considered in the DFT calculations.
Even though the CDW transition opens a partial gap on the Fermi surface and affects the MIR absorptions, the good correspondence between measurements and calculation indicates that the overall band structure is not drastically distorted; this is supported by recent ARPES observations~\cite{Kobata2016, Lei2023}.
Considering the energy size and the possible excitations near the Fermi level, we ascribe the low-energy absorptions (denoted by the orange segment in Fig.~\ref{fig:DFT}c) to the excitations on the Dirac cones along the $\Gamma-Z$ direction in the Brillouin zone (orange arrows in Figs.~\ref{fig:DFT}a and b).
In our measurements, we find this low-energy peak mix with the intraband responses at high $T$s.
In the case of EuAl$_4$, below $T_{CDW}$, the suppressed Drude components give way to the absorptions from Dirac bands, finally resulting in a plateau-like structure in $\sigma_1(\omega)$ with the excitations of CDW gap at 5~K (Fig.\ref{fig:optical spectroscopy}c).
For EuGa$_4$, without the CDW gap, this low-energy peak appears only below $T_N\sim$16~K, when the Drude peak narrows remarkably with the diminishing spin fluctuations (Fig.~\ref{fig:optical spectroscopy}d).
On the other hand, the MIR responses (green segment in Fig.~\ref{fig:DFT}c) can be ascribed to the excitations between bands dominated by Eu $5d$ and Al $3p$ orbitals (green arrows in Figs.~\ref{fig:DFT}a and b).
When Eu's moments are aligned, both experimental and theoretical results evidence that the MIR absorptions are remarkably enhanced, while the change of low-energy peak is minor.
Moreover, in calculated $\sigma_1(\omega)$ (Fig. S6b of SM~\cite{Supplementary}), we notice that only the conductivity from 0.4 to 1.4~eV shows remarkable change in the FM state, indicating that the excitations between Eu $5d$ and Al $3p$ orbitals (green arrows in Figs.~\ref{fig:DFT}a and b) bear primary responsibility for the FM exchange interactions, which is significantly boosted in the forced FM state.

\section{Discussion}
%
In EuAl$_4$, the mechanism of CDW transition remains unclear.
Although the concept of Fermi surface nesting has been proposed, the observation of CDW gaps and band folding remains elusive~\cite{Ramakrishnan2022, Kobata2016}, and the electron-phonon coupling was also proposed to play a decisive role~\cite{Wang2023}.
Here, by comparing with the isostructural and isoelectronic EuGa$_4$, our optical measurement firstly identified the CDW gap with the size around 66~meV.
The CDW transition in EuAl$_4$ partially erodes the Fermi surface, which may come from imperfect nesting between the Dirac-like bands along the $\Gamma-Z$ direction~\cite{Wang2023}.
However, the upshift of the valence band and much weaker electron-phonon coupling~\cite{Wang2023} could be the plausible reason for the absence of CDW order in EuGa$_4$.

In EuM$_4$ family, the itinerant carriers mediate the RKKY AFM interactions between Eu's local moments.
In EuGa$_4$, without CDW order, there forms collinear A-type AFM order, in which the spins lie in the $ab$-plane and perform FM in-plane coupling and AFM coupling along the $c$-axis~\cite{Zhang2021}.
Its magnetism is robust under out-of-plane field up to 7~T~\cite{Shang2021}.
However, in EuAl$_4$, since the CDW transition eliminates part of the Fermi surface, with less carriers, the AFM interactions are suppressed~\cite{Ramakrishnan2022, Takagi2022, Kaneko2021}.
Besides the partial gap, the enhanced MIR absorptions around 0.4~eV after the CDW transition signals promoted FM excitations, which is further bolstered by recent nuclear magnetic resonance and muon spin resonance measurements that observed vigorous out-of-plane FM fluctuations in EuAl$_4$ under zero field~\cite{Zhu2022, Niki2020}.
Thus, by frustrating the AFM interactions and enhancing FM exchange interactions(Fig.~\ref{fig:DFT}(d)), the CDW order in EuAl$_4$ changes the ratio between FM and AFM interactions, intensifying their competition and pushing the system to the alleged quantum critical point, around which the magnetism become unstable~\cite{Gen2023}.
Recently, in layered [MnBi$_2$Te$_4$][Bi$_2$Te$_3$]$_n$ family, thicker nonmagnetic layers between magnetic MnBi$_2$Te$_4$ layers can frustrate the interlayer magnetic coupling and destabilize the magnetism~\cite{Tan2020}.
In EuAl$_4$, we notice that the CDW modulation, which is along the c-axis, offer an alternative approach to destabilize the magnetism.

On the other side, after the CDW transition, the incommensurable lattice distortion breaks the inversion symmetry~\cite{Ramakrishnan2022}, and the growing FM fluctuations at low $T$s will further break the time-reversal symmetry, both of them lift the degeneracy of Dirac bands near the Fermi level, leading to Weyl bands~\cite{Ma2020, Wang2023}.
Because of the partial gap, the Fermi surface will become anisotropic. lift
Therefore, carriers from the Wely bands, which shows the spin-momentum locking, will mediate the anisotropic magnetic interactions, providing the prerequisite for the formation of chiral spin textures~\cite{Gaudet2021,Takagi2022, Kaneko2021}.

\section{Conclusion}
In conclusion, to reveal the interactions among CDW order, topology, and magnetism in EuAl$_4$, we carried out comparative study of isostructure EuAl$_4$ and EuGa$_4$ through optical spectroscopy and the first-principles calculations.
We have found that, by affecting band and lattice structures, the CDW transition in EuAl$_4$ not only modulates the magnetic interactions but also breaks the symmetry.
Due to the intensified competition between AFM and FM interactions and the anisotropic magnetic interactions mediated by carriers from the Weyl bands, the magnetism becomes unstable, and the non-coplanar spin textures such as helical spin order and skyrmions are promoted in EuAl$_4$~\cite{Takagi2022, Shang2021}.
Since the CDW order and intricate magnetism were found in a class of materials with the tetragonal lattice, we propose that the underlying mechanism in EuAl$_4$ is likely to be prevalent across all these materials~\cite{Lei2019, Lei2021}.
Besides the intrinsic anisotropy in polar magnetic topological materials~\cite{Gaudet2021}, the spontaneous symmetry breaking introduced by the CDW transition in EuAl$_4$ provides an alternative way to realize the chiral spin textures in tetragonal lattice.
Since the CDW transition can be further tuned by pressure or charge doping, it offers a new way to tailor the magnetic and topological properties by manipulating the CDW order.

%
%
\section{Acknowledgments}
We thank Artem V. Pronin, Bing Xu, Sheng Li, and Sailong Ju for fruitful discussions, and Gabriele Untereiner for the measurement support. The project was funded by the Deutsche Forschungsgemeinschaft via DR228/51-3. Z.W acknowledges support from the National Natural Science Foundation of China (Grant No. 92065109), the National Key R\&D Program of China (Grant Nos. 2020YFA0308800, 2022YFA1403401), the Beijing Natural Science Foundation (Grant Nos Z190006). R.Yang acknowledges the support from the Alexander von Humboldt foundation.
%
%
%
%
%
%
\section{Author contribution}
P. Z, Z. -W. W and T. S grew the single crystals and carried out the transport measurements. Y. -M. D and R. Y. measured the optical spectroscopy. C. -C. Le performed the first-principles calculations. R.Y. analyzed the data and prepared the manuscript with comments from all authors. R. Y and M. D. supervised this project.
%
\bibliography{EuAl4-bib}

\end{document}